\def\bbbr{{\rm I\!R}} 
\def\Me{Rodr\'{\i}guez}

\documentstyle[twoside,11pt,maxent94,psfig]{article}

\title{Confidence Intervals From One Observation}

\author{C. C. \Me\ \\
	Department of Mathematics and Statistics   \\
	State University of New York at Albany  \\
	E-Mail: carlos@omega.albany.edu  \\
	URL:  http://omega.albany.edu:8008/carlos }    

\begin{document}
\maketitle
\thispagestyle{empty}

\begin{abstract} 
Robert Machol's surprising result, that from a single observation it is possible 
to have finite length confidence intervals for the parameters of location-scale 
models, is re-produced and extended. Two previously unpublished modifications
are included. First, Herbert Robbins nonparametric confidence interval is obtained.
Second, I introduce a technique for obtaining confidence intervals for the
scale parameter of finite length in the logarithmic metric. 
\end{abstract}

\section{Introduction}
Let $x$ be an observation from a $N(\mu,\sigma^2)$ population with unknown
parameters. The following statement belongs to the folklore of Statistical
Science:
\smallskip\noindent
{\leftskip 1in\rightskip 1in
{\it From a single observation $x$ we can not gain information about
the variability in the population. Thus, finite length confidence intervals 
for $\mu$ and/or $\sigma$ are impossible even in principle}.
}

This is not correct. For example $x\pm 5\cdot|x|$ will cover $\mu$
at least 90\% of the time and $(0,17|x|)$ will cover $\sigma$ at least
95\% of the time. If you don't believe it check it with your PC!

I first heard about this some years ago from Herbert Robbins. According to Robbins, this
phenomenon was discovered by an electrical engineer in the 60's (Robert Machol
{\it IEEE Trans. Info. Theor.}, 1964) but it is still relatively unknown to 
statisticians.

I show Machol's idea below. The intervals for $\mu$ in the parametric
case are due to him. The nonparametric improvement is due to Robbins and the
intervals on $\sigma$ are mine.

\section{Confidence Intervals for $\mu$, Parametric Case}

Consider the following problem. Given a single observation from a r.v.
$$ X \leadsto {1 \over \sigma}\cdot f({{x-\mu}\over\sigma}),\  \mu\epsilon\bbbr,
\  \sigma > 0\   {\rm unknown}, $$
with $f$ a {\it known} density symmetric about zero. Find a finite length
$100\cdot(1-\beta)\% $ CI for $\mu$. \\
\smallskip\noindent {\bf Machol's answer}: Consider the event
$$ A = [ |X - \mu| > t |X - a| ] $$
where $a\epsilon\bbbr$ is an arbitrary constant and $t > 1$ is given. We have
$$ A = [ |Y| > t |Y - \alpha| ] $$
where 
$$ Y = {{X - \mu}\over{\sigma}} \leadsto {f(y)} {\rm\  and\ \ } 
\alpha = {{a - \mu}\over{\sigma}}\ \epsilon\bbbr.  $$
The event $A$ corresponds to the shaded piece in Fig. 1. Thus,
\begin{figure}
\centerline{
\psfig{figure=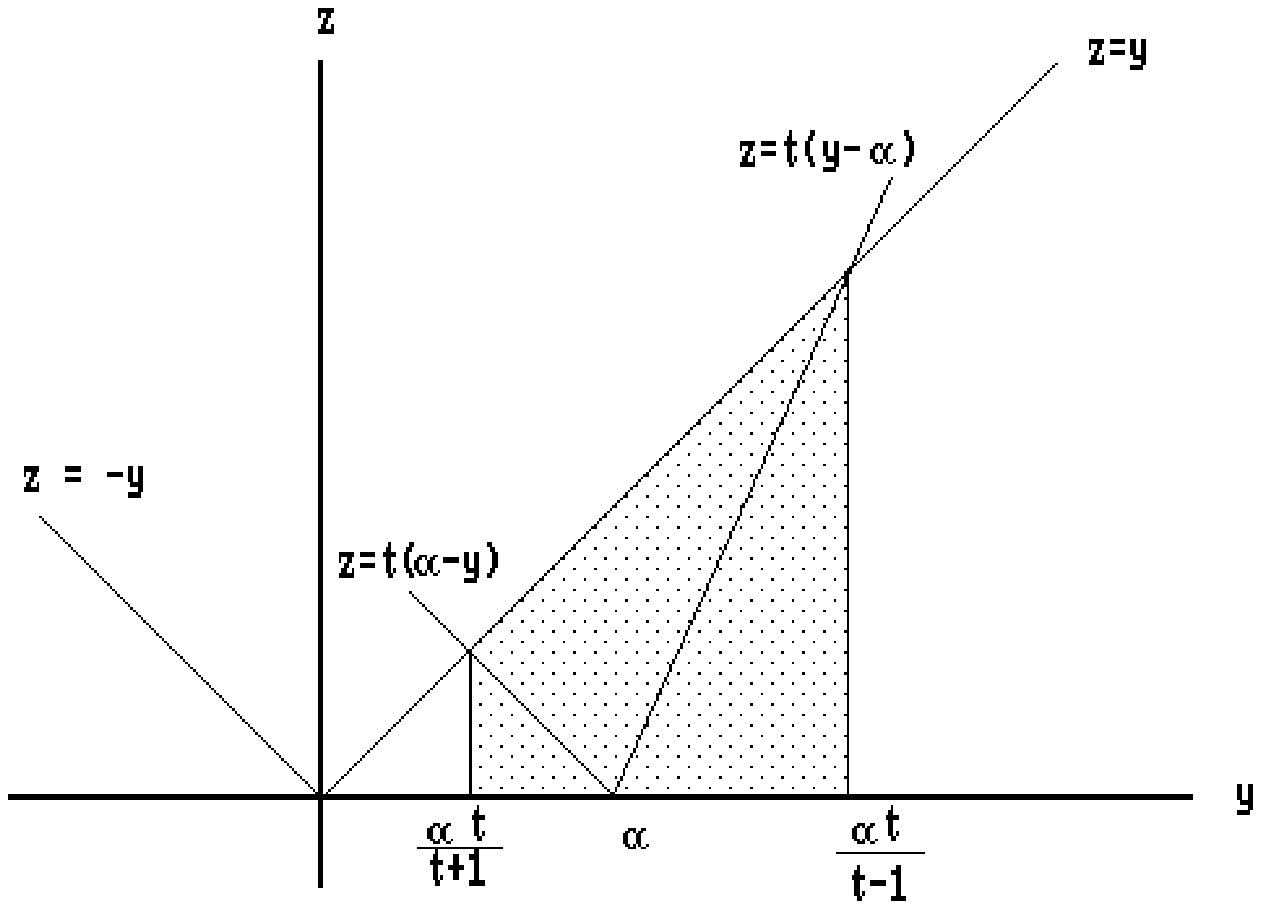,width=3.5in}}
\centerline{Fig. ~1. Illustration of event A}
\end{figure}
$$ P(A) = P[\ |Y| > t |Y - \alpha|\ ] = \left| {\int_{{\alpha t}\over{t+1}}
^{{\alpha t}\over{t-1}} f(y) dy }\right| = \beta(\alpha,t) $$
and 
$$ P(A) \le \beta^*(t) = \sup_{\alpha\epsilon\bbbr} \beta(\alpha,t).  $$
Therefore
$$ P[\ X - t|X - a| \le \mu \le X + t|X - a|\ ] = P(A^c) \ge 1 - \beta^*(t) $$
Hence, provided that $\beta^*(t) \to 0$ as $t \to \infty$ the interval
$X \pm t|X-a|$ can be made to have any pre-specified confidence. \\
\smallskip\noindent{\bf Example}: Take $f(y) = \phi(y) \equiv 
\  {\rm pdf\ of\ } N(0,1)$. From the symmetry of $\phi$ about 
zero we can write
$$ \beta(-\alpha,t) = \left| {\int_{-{{\alpha t}\over{t+1}}}^{-{
{\alpha t}\over{t-1}}} \phi(z) dz }\right| = \beta(\alpha,t) $$
Thus,
$$ \beta^*(t) = \sup_{\alpha > 0} \beta(\alpha,t). $$
For $\alpha > 0$ we have,
$$ {{{\partial\beta}\over{\partial\alpha}}(\alpha,t)} = 
{ {t\over{t-1}}\phi\left({{\alpha t}\over{t-1}}\right)\  -\ 
  {t\over{t+1}}\phi\left({{\alpha t}\over{t+1}}\right) }\ = 0, $$
so that
$$ \exp\left[{ {1\over 2}\left({{\alpha t}\over{t+1}}\right)^2\ - {1\over 2}
\left({{\alpha t}\over{t+1}}\right)^2}\right]\ = {{t+1}\over{t-1}} $$
and taking logs we obtain
$$ {{{\alpha^2 t^2}\over{(t^2-1)^2}}[(t^2+2t+1)-(t^2-2t+1)]} = {2 \log\left(
{{t+1}\over{t-1}}\right)} $$
from where
$$ \alpha^* = { {{t^2-1}\over t} \sqrt{ {1\over{2t}} {\log\left({{t+1}
\over{t-1}}\right)} } } $$
and 
$$ \beta^*(t) = \int_{(t-1)\sqrt{{1\over{2t}}\log\left({{t+1}
\over{t-1}}\right)}}^{(t+1)\sqrt{{1\over{2t}}\log\left({{t+1}
\over{t-1}}\right)}} \phi(y) dy $$
with a calculator and a normal table we find that for $t = 5$ then $\alpha^* =
1.0796,\ \beta^* = .1$ and the confidence is 90\% for $x\pm 5|x|$. Other 
intervals could be computed in a similar way. In fact this shows that
$$ P[\ X - 5|X - a| \le\ \mu\ \le\ X + 5|X - a|\ ]\ >\ .90 $$
for all $a\epsilon\bbbr,\ \mu\epsilon\bbbr\ {\rm and}\ \sigma > 0$.
\par The best $a$ is the one that produces the shortest expected length. But,
{\it length} $= L = 2t|X - a|$ and
$$ E(L) = 2 t E(|X - a|) \propto\ E(|X - a|) $$
so that the best $a = a^*$ should minimize $E(|X - a|)$ i.e. $a^*$ must be
the {\bf median} of $X$ and since $X$ is symmetric about $\mu$ we have
$a^* = \mu$. Hence, the best $a$ is our best a priori guess for $\mu$. This
looks like {\it Bayesianism} sneaking in classical confidence intervals!.
\par The arbitrariness of $a$ in the statement "$x\pm t|x -a|$ is a 
$(1 - \beta^*(t))100\%$ CI for $\mu$" reminds me of the Stein shrinking
phenomenon. Perhaps this is part of the reason why Robbins got interested
in it. Recall that Robbins' Empirical Bayesianism produces Stein's estimators
as a special case.
\section{Confidence Intervals for $\mu$, Non-parametric Case}
Let $\Im$ be the class of all unimodal, symmetric about zero densities.
Given a single observation of $X$ with $X \leadsto f(x - \mu)$ where both
$f \epsilon\Im\ {\rm and}\ \mu\epsilon\bbbr\ $ are {\bf unknown}, find a $100(1-
\beta)\%$ CI for $\mu$ of finite length. \\
\smallskip\noindent{\bf Robbins' Answer}: Consider first the following 
simple lemma: \\
\smallskip\noindent{\bf Lemma}: If $f \downarrow$ in $(0,+\infty)$ then
$$ l(x)\ =\ {1\over{b-x}} {\int_x}^b f(y)\ dy\  \downarrow {\rm in\ } (0,b) $$
{\bf proof}: This is  obvious from the picture (see Fig. 2.), since 
$l(x)$ denotes the mean value of $f$ on $(x,b)$.
\begin{figure}
\centerline{
\psfig{figure=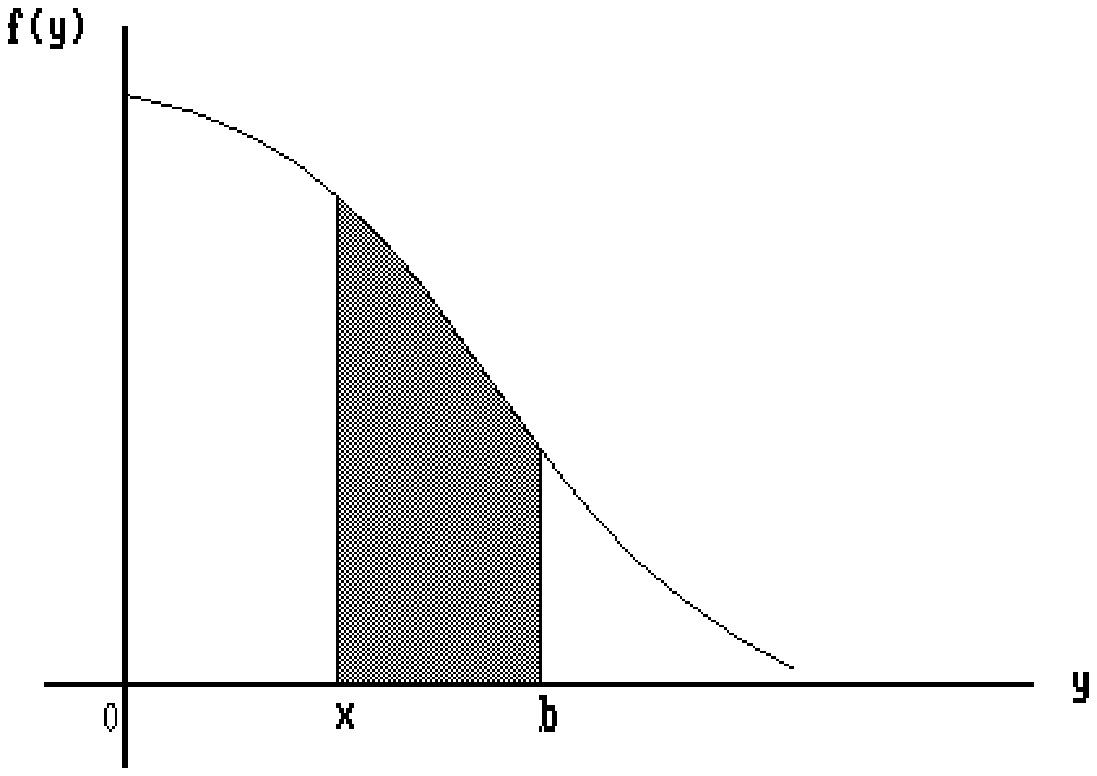,width=3.5in}}
\centerline{Fig. ~2. The mean value of $f(y)$ decreases when x approaches b}
\end{figure}
Of course the algebra gives the same answer. Notice that
$$ {l(x)}\ \le\ { {1\over{b - x}} f(x)\ (b - x) }\ =\ {f(x)}. $$
Thus, differentiating both sides of the equation
$$ {(b - x)\ l(x)}\ =\ { {\int_x}^b\ f(y)\ dy }, $$
we obtain
$$ {l'(x)}\ =\ { {1\over{b - x}} [ l(x) - f(x) ]}\ \le\ 0 $$
i.e. $l(x)$ decreases in $(0,b) \bullet$
\par
\medskip\noindent Consider as before the event
$$ A = {[\ |X - \mu| > t|X - a|\ ]}\ {\rm for}\ t > 1\ {\rm and}\ 
a\epsilon\bbbr. $$
Then, if $Y = X - \mu\ $, we have
$$ {P(A)}\ =\ {P[\ |Y| > t |Y - \alpha|\ ]}\ {\rm with}\ \alpha = 
a - \mu\ \epsilon\bbbr. $$
$$ P(A)\ =\ {\beta(\alpha,t)} = \beta(-\alpha,t)\ {\rm since}\ 
f \epsilon\Im. $$
But now applying the Lemma for $x = {{\alpha t} / (t+1)} > 0$ and $b = 
{\alpha t/ (t-1)}$ we obtain
$$ {l(x)} = { P(A)\over{\alpha t \left({{1\over{t-1}} - {1\over{t+1}}}\right)}
}\le\ {l(0)}\ =\ { {{t-1}\over{\alpha t}} {\int_0}^{{\alpha t}\over{t-1}} f(y)\ 
dy }\ \le\ { {t-1}\over{2\alpha t} }. $$
Hence,
$$ P(A)\ \le\ {1\over{t + 1}} \  {\rm for\ all\ } \alpha\epsilon\bbbr\ 
{\rm and\ } f \epsilon\Im.  $$
Therefore
$$ P[\ X - t|X - a|\ \le\ \mu\ \le\ X + t|X - a|\ ]\ \ge\ {1 - {1\over{1+t}}} $$
holds for all $a\epsilon\bbbr,\ \mu\epsilon\bbbr$, and $f\epsilon\Im.$ \\
\smallskip\noindent {\bf Example}: For $t = 9$, we have $1 - 1/(1+t) = .9$, 
and $x\pm 9|x-a|\ $ will cover $\mu$ at least $90\%$ of the time even if we
are uncertain about $f\epsilon\Im$. This suggests the following game: Each 
time you pick up a function $f$ in $\Im$ in any way you want i.e. 
deterministically or stochastically with some distribution. Then you choose 
$\mu\epsilon\bbbr\ $ also in an arbitrary way i.e. each $\mu\ $ every time or
following a pre-specified sequence, or generate them with a distribution
changing the distribution each time etc... Then use the computer to show me
$x\leadsto f(x-\mu)\ $. I win \$1 if $x\pm 9|x|\ $ covers your $\mu\ $ and you
win \$5 if it doesn't. Do you want to play a couple of hundred times?

\section{Confidence Intervals for $\sigma$}

We consider now the estimation of the scale parameter from a single observation.
It should be noticed that the only interesting confidence intervals are those
of finite length. Thus, $(0,\infty)$ is a $100\%$ confidence interval but
useless. 

The natural, invariant under re-parameterizations, measure of length for
a confidence interval $(a,b)$ for a scale parameter is not just $b-a$
but proportional to the difference in the logarithmic scale,
i.e. $\log b - \log a$. This follows by recalling the fact that the square of the
element of length, on the hypothesis space of the location-scale
model, along a line of constant scale is given by:

\[ ds^{2} = g_{\sigma \sigma} (d\sigma)^{2} \]
where $g_{\sigma \sigma}$ is the Fisher information amount at $\sigma$ given by:

\[ g_{\sigma \sigma} = {{k-1}\over\sigma^{2}} \]
with

\[ k = 4 \int_{-\infty}^{\infty} y^{2} \left( \psi'(y) \right)^{2} dy \]
and $\psi^{2} = f$ in the notation of the proposition below. Hence, the
geodesic distance from the probability distribution with scale ``$a$''
to the probability distribution with scale ``$b$'' is obtained by
integrating the element of length and therefore proportional to the difference
in the log scale as noted above. The reader unfamiliar with the geometry of
hypothesis spaces may use the expression of the Kullback number between the
gaussian with mean zero and standard deviation ``$a$'' and the gaussian with
mean zero and standard deviation ``$b$'' as an approximation to the geodesic
distance, to convince him/herself of the logarithmic nature of this length.

It is therefore necessary to consider confidence intervals with
non-zero lower bounds, since $\sigma = 0$ is in fact a line at
infinity. I show below that it is possible to have finite length
confidence intervals for the scale parameter from a single
observation, but only if we rule out a priori from the hypothesis space
a bit more than the line $\sigma = 0$. It is this interplay between
geometry, classical inference and bayesianism that I find appealing in
this problem.

{\bf Proposition}: Let $f$ be a pdf symmetric about 0 and differentiable
everywhere. Let $F$ be the associated cdf. Let $0 < t_1 < t_2 \le \infty\ $
with $f'(t_1) > f'(t_2)\ $ and define
$$ {G(\alpha,t_1,t_2)}\ =\ F(\alpha - t_1) + F(\alpha + t_2) - F(\alpha - 
t_2) - F(\alpha + t_1). $$
Let $M > 0,\ a\epsilon\bbbr,\ \mu\epsilon\bbbr,\ \sigma > 0\ $ be given numbers.
Then if 
$$|\mu - a|\ \le\ \sigma M\ {\rm and\ } X\leadsto\ {{1\over\sigma} f\left({
{x-\mu}\over\sigma}\right)},$$ 
we have
$$ P\left[{\ {{|X - a|}\over{t_2}}\ \le\ \sigma\ \le\ {{|X - a|}\over
{t_1}}\ }\right]\ \ge\ 2 [F(t_2) - F(t_1)]\ I[M\le M^*]\ + $$
$$\ \ \ \ \ \ \ \ \ \ I[M > M^*]\ \inf_{0 < \alpha < M} 
\bigl\{G(\alpha,t_1,t_2)\bigr\}. $$
Where $M^* = \min\ \{ \alpha > 0 :\ G(\alpha,t_1,t_2)\ =\ G(0,t_1,t_2)\ \}. $
If $f\equiv N(0,1)\ $ (or any other pdf with similar tails) and excellent
approximation is 
$$ M^*\ =\ t_2\ +\ F^{-1}(2F(t_1) - 1) $$
{\bf Proof}: Consider the event
$$ A\ =\ { \left[{ {{|X - a|}\over{t_2}}\ \le\ \sigma\ \le\ {{|X - a|}\over
{t_1}} }\right] }. $$
Let
$$ Y\ =\ { {X - \mu}\over\sigma } \leadsto\ f(y). $$
Then by adding and subtracting $\mu$ inside the absolute values and dividing
through by $\sigma$ we obtain
$$ A\ =\ { [t_1\ \le\ |Y - \alpha|\ \le\ t_2] } $$
where $\alpha = (a-\mu)/\sigma\ $ is such that $|\alpha|\ \le\ M$. Notice that
the y's satisfying the inequalities that define the event $A$ correspond to
the shaded region in Fig. 3.
\begin{figure}
\centerline{
\psfig{figure=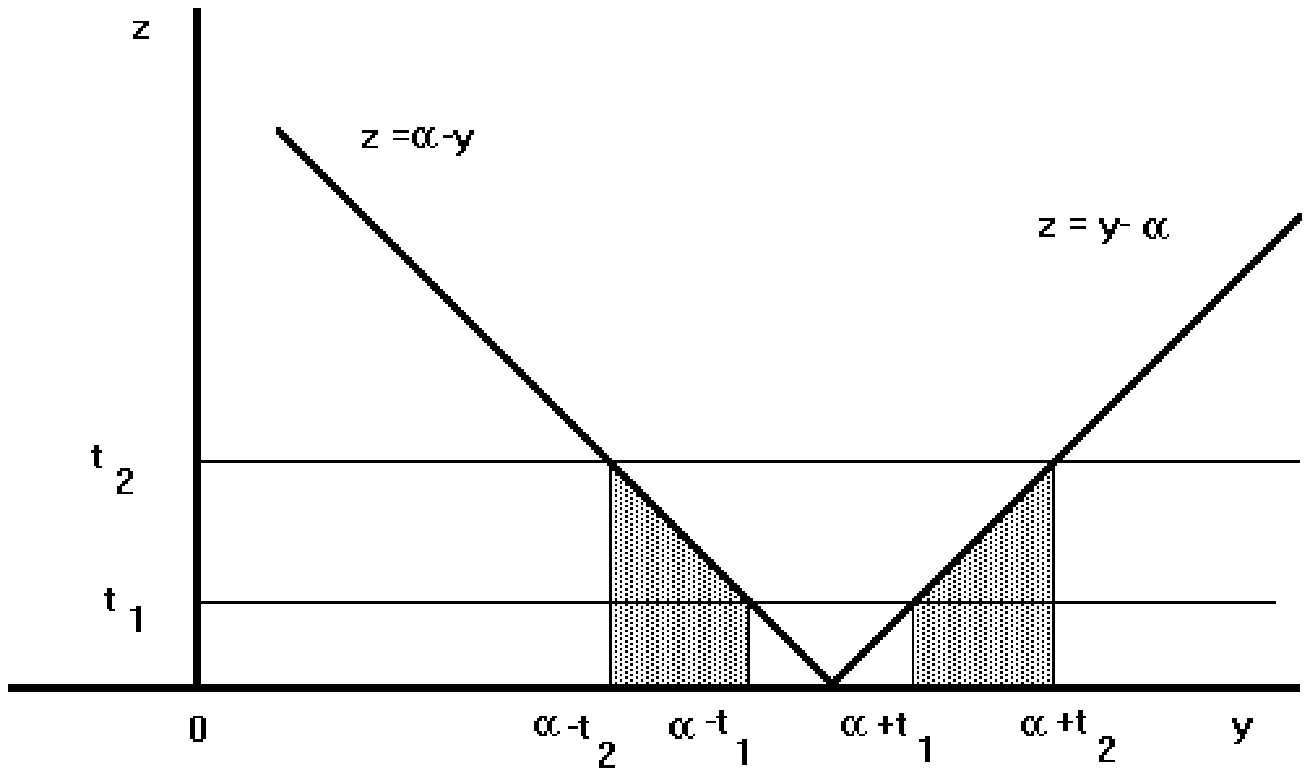,width=3.5in}}
\centerline{Fig. ~3. Illustration of event $A$}
\end{figure}

Hence,
$$ P(A)\ =\ { \int_{\alpha - t_2}^{\alpha - t_1} f(y) dy\ +\ 
\int_{\alpha + t_1}^{\alpha + t_2} f(y) dy\ =\ G(\alpha,t_1,t_2) } $$
Notice that for given values $t_1$ and $t_2$ the function $G$, as a function 
of $\alpha$ is twice differentiable and symmetric about zero with a local
minimum at $\alpha = 0$. Since, using the fact that $f(y) = f(-y)$ we have
$${ \left.{{\partial G}\over{\partial \alpha}}\right|_{\alpha = 0}\ } = {
\ \left.{\left[{f(\alpha - t_1) - f(\alpha - t_2) + f(\alpha + t_2) - 
f(\alpha + t_1)}\right]}\right|_{\alpha = 0}\ } ={\ 0} $$
and also
$$ \left.{{\partial^2 G}\over{\partial\alpha^2}}\right|_{\alpha = 0}\ =\ 
f'(-t_1) - f'(-t_2) + f'(t_2) - f'(t_1) $$
$$ =\ 2\ (f'(t_1) - f'(t_2))\ >\ 0\ \ \ \ \ \ \ $$
Thus,
$$ P(A)\ \ge\ G(0,t_1,t_2)\ =\ 2 [F(t_2) - F(t_1)] $$
provided that $|\alpha| \le M^*$ i.e. if $M \le M^*$. The picture (see Fig. 4.)
illustrates the situation.
\begin{figure}
\centerline{
\psfig{figure=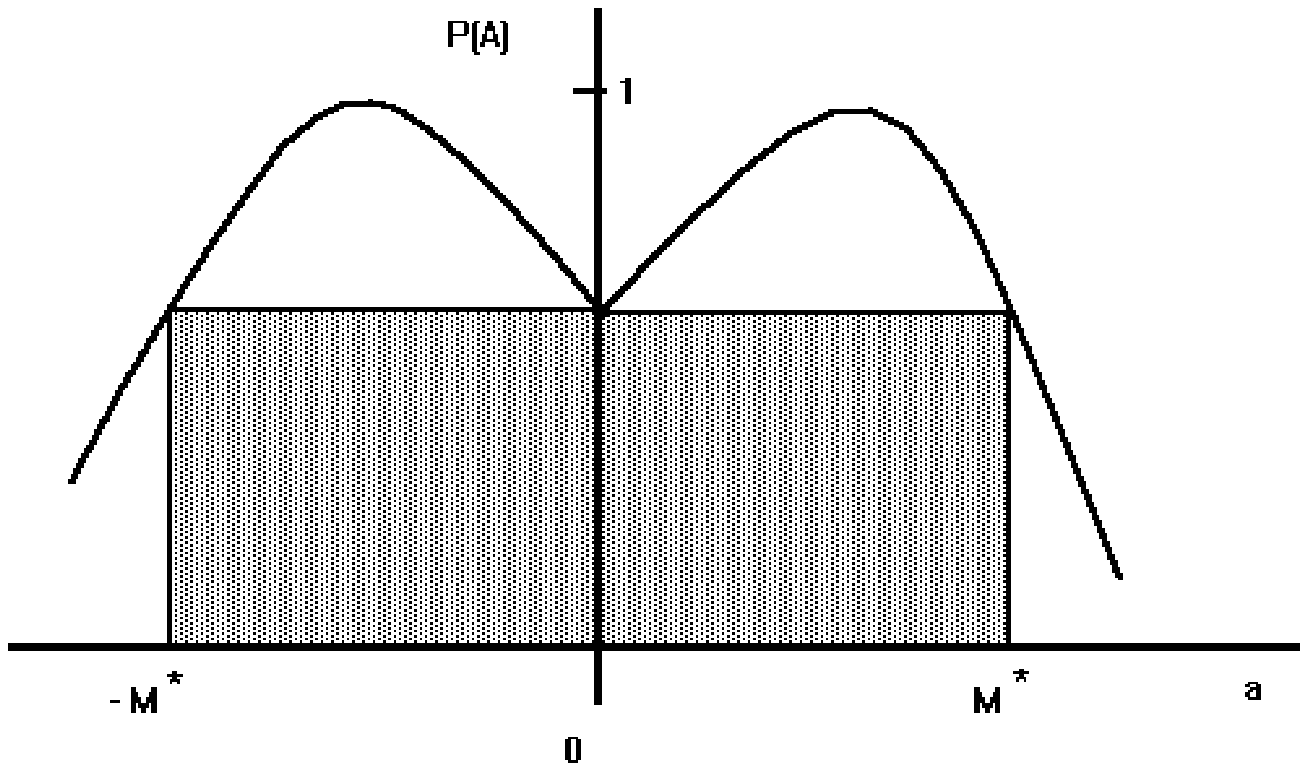,width=3.5in}}
\centerline{Fig. ~4. Illustration of the event $A$}
\end{figure}

\par In the gaussian case, to obtain reasonable confidences we must have
$t_1 < 1$ and $t_2 > 3$. Hence, $ F(\alpha - t_1)\ \approx\ F(\alpha + t_1)
\ \approx\ F(\alpha) $ and $ F(\alpha + t_2)\ \approx\ 1$. From where
$$ G(\alpha,t_1,t_2)\ \approx\ 1 - F(\alpha - t_2)\ \equiv\ 2 [1 - F(t_1)]
\ \approx\ G(0,t_1,t_2) $$
and the approximation for $M^*$ is obtained by solving the central identity
for $\alpha \bullet$	\\
\medskip\noindent{\bf Remarks}:
\par 1) Notice that the lower bound of the confidence interval, i.e. $|x - a|
/ t_2$, is positive only if $M < \infty$ i.e. if we know a priori that
$|\mu - a| \le \sigma M < \infty$.
\par 2) When $t_2 \to \infty$ then $M^* \to \infty$ and with no prior 
knowledge ( i.e. $|\mu - a| < \infty$ ) we still have
$$ P \left({0 \le \sigma \le {{|X - a|}\over{t_1}}}\right)\ \ge\ 
2 (1 - F(t_1)). $$
\par 3) The value of $t_2$ is related to the amount of prior information.
 The larger $t_2$ the weaker the prior information necessary to assume
the desire confidence. On the other hand $t_1$ controls the confidence
associated to the interval. These remarks are illustrated with examples. \\
\medskip\noindent{\bf Examples}: Let $x$ be a single observation from
a gaussian with unknown mean $\mu$ and unknown variance $\sigma^2$. Then
$90\%$ CIs for $\sigma$ are:
\par $(0,8|x|)\ \ \ $ valid always 
\bigskip\par $\left({{{|x|}\over{4}}, 8 |x| }\right) \ \ $ valid if $|\mu|\le 2.7\sigma$
\bigskip\par $\left({{{|x|}\over{8}}, 8 |x| }\right) \ \ $ valid if $|\mu|\le 6.7\sigma$

$95\%$ CIs are:
\bigskip\par $\left({{{|x|}\over{5}}, 17|x| }\right) \ \ $ valid if $|\mu|\le 3.3\sigma$
\bigskip\par $\left({{{|x|}\over{50}},17|x| }\right) \ \ $ valid if $|\mu|\le 48\sigma$
\bigskip\par $\left({0 , 17 |x| }\right) \ \ \ \ $   valid always.

$99\%$ CIs are:
\bigskip\par $\left({{{|x|}\over{5}},70|x| }\right) \ \ $ valid if $|\mu|\le 2.7\sigma$
\bigskip\par $\left({{{|x|}\over{10^3}},70|x|}\right) \ \ $valid if $|\mu|\le 997\sigma$
\bigskip\par $(0,70|x|)\ \ \ $ valid always.

\section*{Almost Real Example}
I'll try to show that the required
prior knowledge necessary to have non-zero lower bounds for the CIs is in fact
often available. Suppose that we want to measure the length of the desk in my
office with a regular meter graduated in centimeters. Let $x$ be the result of
a single measurement and let $\mu$ be the {\it true} length of my desk. Then
$$ x = \mu + \varepsilon {\rm\ with\ } \varepsilon\ \leadsto\ N(0,\sigma^2) $$
is a reasonable and very popular assumption. Now, even before I make the 
measurement I can write with {\it all} confidence that for my desk 
$\mu = 2 \pm 1m$ i.e. $|\mu - 2| \le 1$. With the meter graduated in
centimeters I will be guessing the middle line between centimeters so I can
be sure that $x = \mu \pm$ at least $1\over 4$ of a centimeter. Thus,
$$ 3 \sigma \ge {1 \over 400}. $$
Therefore I can be absolutely sure that
$$ |\mu - 2|\ \le\ 1200 \sigma. $$
Hence,
$$ \left({{{|x - 2|}\over{1500}}, 70 |x - 2|}\right) $$
will be a $99\%$ CI for $\sigma$.

\end{document}